\documentclass{moriond}

\def\be{\begin{equation}}
\def\ee{\end{equation}}
\def\bea{\begin{eqnarray}}
\def\eea{\end{eqnarray}}

\newcommand{\Photo}{}

\begin{document}
\vspace*{4cm}
\title{GLOBAL FIT ON COLOURED SCALARS INCLUDING THE LAST W-BOSON MASS MEASUREMENT}

\author{ V. MIRALLES$^{1,}$\,\footnote{Speaker and corresponding author: victor.miralles@roma1.infn.it}, O. EBERHARDT$^2$, H. GISBERT$^3$,  A. PICH$^2$, J. RUIZ-VIDAL$^2$ }

\address{$^{1}$INFN, Sezione di Roma, Piazzale A. Moro 2, I-00185 Roma, Italy\\
$^{2}$IFIC, Universitat de Val\`encia-CSIC, Apt. Correus 22085, E-46071 Val\`encia, Spain\\
$^{3}$Fakult\"{a}t f\"{u}r Physik, TU Dortmund, Otto-Hahn-Str. 4, D-44221 Dortmund, Germany
}

\maketitle\abstracts{
We consider a simple extension of the electroweak theory, incorporating one $SU(2)_L$ doublet of
colour-octet scalars with Yukawa couplings satisfying the principle of minimal flavour violation.
Using the HEPfit package, we perform a global fit to the available data, including all relevant theoretical constraints, and extract the current bounds on the model parameters. We also calculate the contributions of the additional coloured scalars to the electric dipole moment of the neutron and set limits on its parameter space using the current experimental information. Finally, we study the effect that the last measurement of the $W$-boson mass has in our fit and show how the colour-octet scalar extension struggles to accommodate the new measurement.
}

\section{Introduction}

Besides the enormous success that the Standard Model (SM) has achieved, anticipating a wide range of phenomena and predicting experimental outcomes, there are some phenomena that need further understanding. For the moment there is (apparently) only one fundamental scalar particle, the Higgs boson, but there is no evidence of it being the only one. During the past years many extensions of the scalar sector of the SM have been proposed and in this work we will focus on one of them that extends the scalar sector with a $SU(2)_L$ doublet which is an octet of $SU(3)_C$. This model was first proposed by Manohar and Wise (MW)~\cite{Manohar:2006ga}, from whom receives its name, and its initial motivation was that this scalar representation is one of the few that can be compatible with the principle of minimal flavour violation (MFV). This type of scalars, with masses of few TeVs, also emerge naturally in many SU(4), SU(5) and SO(10) grand unification theories (GUTs), see \textit{e.g.} Refs.~\citelow{Dorsner:2007fy,FileviezPerez:2013zmv,Perez:2016qbo}. 

We will first perform a global fit~\cite{Eberhardt:2021ebh} of the CP-conserving MW model, using the \texttt{HEPfit} package~\cite{deBlas:2019okz}. After that we discuss the CP-violating case and predict~\cite{Gisbert:2021htg} the contribution of the coloured scalars to the neutron electric dipole moment (EDM).
Finally, and given the large relevance of the new measurement of the $W$-boson mass by the CDF collaboration \cite{CDF:2022hxs}, we show how the model struggles to accommodate this measurement, studying its effect in the oblique parameters \cite{Peskin:1990zt,Peskin:1991sw}.\footnote{ At the time of giving the talk the measurement of CDF was still not public. }

\section{The colour-octet scalar model}

The scalar sector of the SM is extended with one electroweak doublet of colour-octet scalar fields with hypercharge $Y=\frac{1}{2}$. Since it has colour charge, the new scalar multiplet does not mix with the SM Higgs doublet. As colour must be conserved, the coloured particles cannot acquire a vacuum expectation value (VEV). Therefore, only the SM Higgs boson will acquire a VEV which will minimise the most general potential that can be built with this scalar sector:
\begin{equation}
\begin{array}{ll}
 V_{\rm{\tiny{MW}}}&
  = m_\Phi^2\Phi^\dagger\Phi + \frac12 \lambda \left(\Phi^\dagger\Phi \right)^2 + 2 m_S^2 {\rm Tr}\left(S^{\dagger i} S^{\phantom{\dagger}}_i\right)+ \mu_1 {\rm Tr}\left(S^{\dagger i} S^{\phantom{\dagger}}_i S^{\dagger j} S^{\phantom{\dagger}}_j\right) 
  + \mu_2 {\rm Tr}\left(S^{\dagger i} S^{\phantom{\dagger}}_j S^{\dagger j} S^{\phantom{\dagger}}_i\right)  \nonumber  
  \\
& \phantom{{}={}}
 + \mu_3 {\rm Tr}\left(S^{\dagger i} S^{\phantom{\dagger}}_i\right) {\rm Tr}\left(S^{\dagger j} S^{\phantom{\dagger}}_j\right)
  + \mu_4 {\rm Tr}\left(S^{\dagger i} S^{\phantom{\dagger}}_j\right) {\rm Tr}\left(S^{\dagger j} S^{\phantom{\dagger}}_i\right) 
  + \mu_5 {\rm Tr}\left(S^{\phantom{\dagger}}_i S^{\phantom{\dagger}}_j\right) {\rm Tr} \left(S^{\dagger i} S^{\dagger j}\right) 
   \nonumber \\
&\phantom{{}={}}
  + \mu_6 {\rm Tr}\left(S^{\phantom{\dagger}}_i S^{\phantom{\dagger}}_j S^{\dagger j} S^{\dagger i}\right)+ \nu_1 \Phi^{\dagger i}\Phi_{i} {\rm Tr}\left(S^{\dagger j} S^{\phantom{\dagger}}_j\right) 
  + \nu_2 \Phi^{\dagger i}\Phi_{j} {\rm Tr}\left(S^{\dagger j} S^{\phantom{\dagger}}_i\right) \nonumber \\
 &\phantom{{}={}}
 +\left[ \nu_3 \Phi^{\dagger i}\Phi^{\dagger j} {\rm Tr}\left(S^{\phantom{\dagger}}_i S^{\phantom{\dagger}}_j\right) 
           +\nu_4 \Phi^{\dagger i} {\rm Tr}\left(S^{\dagger j} S^{\phantom{\dagger}}_j S^{\phantom{\dagger}}_i\right) 
           +\nu_5 \Phi^{\dagger i} {\rm Tr}\left(S^{\dagger j} S^{\phantom{\dagger}}_i S^{\phantom{\dagger}}_j\right) 
           + {\rm h.c.}\right] , 
\label{eq:genpot}
\end{array}
\end{equation}
where $\Phi = (\phi^+,\phi^0)^T$ is the usual SM doublet, 
the traces are taken in colour space, and  $i$ and $j$ denote $SU(2)_L$ indices. The additional $(8,2)_{1/2}$ scalar fields
$S^A =(S^{A,+},S^{A,0})^T$ are contained in the multiplet $S=S^AT^A$ with $T^A$ the generators of the $SU(3)_C$ group.

The VEV will also generate an splitting on the masses of the physical particles
\begin{equation}
    m^2_{S^\pm}=m^2_S+\nu_1 \,\frac{v^2}{4}\, ,
    \qquad \qquad
    m^2_{R,I}=m^2_S+(\nu_1+\nu_2\pm 2\, \nu_3)\, \frac{v^2}{4}\, .
    \label{eq:msplit}
\end{equation}

Finally, and since we are assuming MFV, the Yukawa matrices that generate the interaction of the coloured scalars with the quarks will be proportional to the SM Yukawas:
\begin{equation}
 {\cal L}_Y \supset -\sum^3_{i,j=1} \left[\eta_D Y^d_{ij}\,\bar{Q}_{L_i}S d_{R_j}+\eta_U Y^u_{ij}\,\bar Q_{L_i}\tilde{S}u_{R_j} + \rm{h.c.}\right] .
 \label{eq:LY}
\end{equation}

\section{CP-conserving case: Global fit}

In this section we will show the constraints that we have found on the parameter space of this model, including both theoretical assumptions and experimental data. This analysis has been performed in the CP-conserving limit, in order to reduce the total number of parameters. Therefore, all the parameters of the potential will be considered as real, as well as the Yukawa proportionality coupling constants, $\eta_U$ and $\eta_D$. However, having complex Yukawa couplings can have a strong impact on the EDM of the neutron, as we will see in Sec.~\ref{sec:EDMs}. The experimental information used in the current section will be the same as in Ref.~\citelow{Eberhardt:2021ebh} and we leave the discussion of the new measurement of the W-boson mass by CDF for Sec.~\ref{sec:CDFmeasurement}. In order to perform the fits we have used the $\texttt{HEPfit}$ package, which has been proven to be extremely useful for fitting many new physics models, as well as the SM \cite{deBlas:2021wap,deBlas:2022hdk} and even effective field theories \cite{Durieux:2019rbz,Miralles:2021dyw,Durieux:2022cvf}. Using this tool, we will do a bayesian fit for which we have used the following uniformly distributed priors
\begin{table}[h!]
\begin{center}
\begin{tabular}{|c|c|c|c|c|c|}
\hline
Parameters&$\textcolor{black}{m_S^2}$&$\textcolor{black}{\nu_n}$&$\textcolor{black}{\mu_n}$&$\textcolor{black}{\eta_U}$&$\textcolor{black}{\eta_D}$\\
\hline
Priors &$\textcolor{black}{(0.4^2, 1.5^2)}$ TeV$\textcolor{black}{^2}$&(-10, 10)&(-10, 10)&(-5, 5)&(-20, 20)\\
\hline
\end{tabular}
\end{center}
\end{table}

\subsection{Theoretical assumptions and experimental observables}

The theoretical assumptions that we have considered are  renormalisation group (RG) stability, perturbative unitarity and perturbative behaviour of the quantum corrections. In order to guarantee RG stability we have imposed the absence of Landau poles and that the potential should be bounded from below up to a desired scale, $\mu_{st}$, set to 3 or 5 TeV. Perturbative unitarity means that the probability of the two-to-two scattering should be smaller than one. For imposing these conditions we have made use of the expressions given in Ref.~\citelow{He:2013tla}  at leading order (LO). These expressions are calculated in the large $s$ approximation so we imposed them for scales higher than $\mu_u=1.5$ TeV, above the maximum value of the mass region covered here. The next-to-LO (NLO) and NLO+ expressions were calculated previously \cite{Cheng:2018mkc} following the procedure of Ref.~\citelow{Murphy:2017ojk}.  Note that the NLO+ expressions contain some terms at next-to-NLO, for more details we refer to Ref.~\citelow{Murphy:2017ojk}.

With respect to the experimental constraints we have made used of direct searches  \cite{Hayreter:2017wra,Miralles:2019uzg}, Higgs signal strengths (HSS) \cite{Cheng:2016tlc}, meson mixing \cite{Cheng:2015lsa}, the $B_s\rightarrow \mu^+ \mu^-$ decay \cite{Cheng:2015lsa}, the oblique parameters (S, T, U) \cite{Burgess:2009wm} and $R_b$ \cite{Degrassi:2010ne}.

\subsection{Results}

\begin{figure}[h!]
\begin{minipage}{0.45\linewidth}
    \centering
    \includegraphics[scale=0.27]{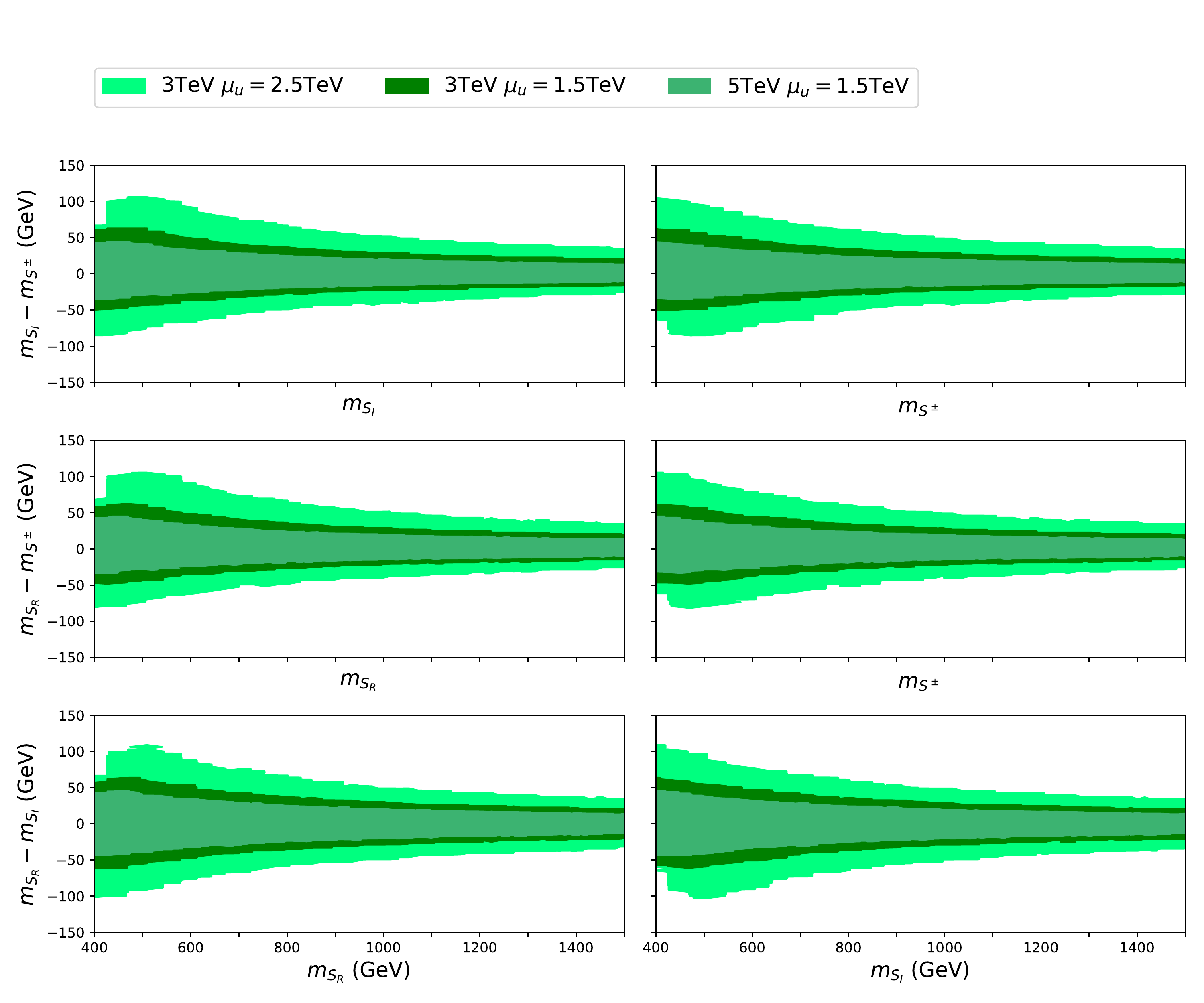}
\end{minipage}
\begin{minipage}{0.45\linewidth}
    \centering
    \includegraphics[scale=0.27]{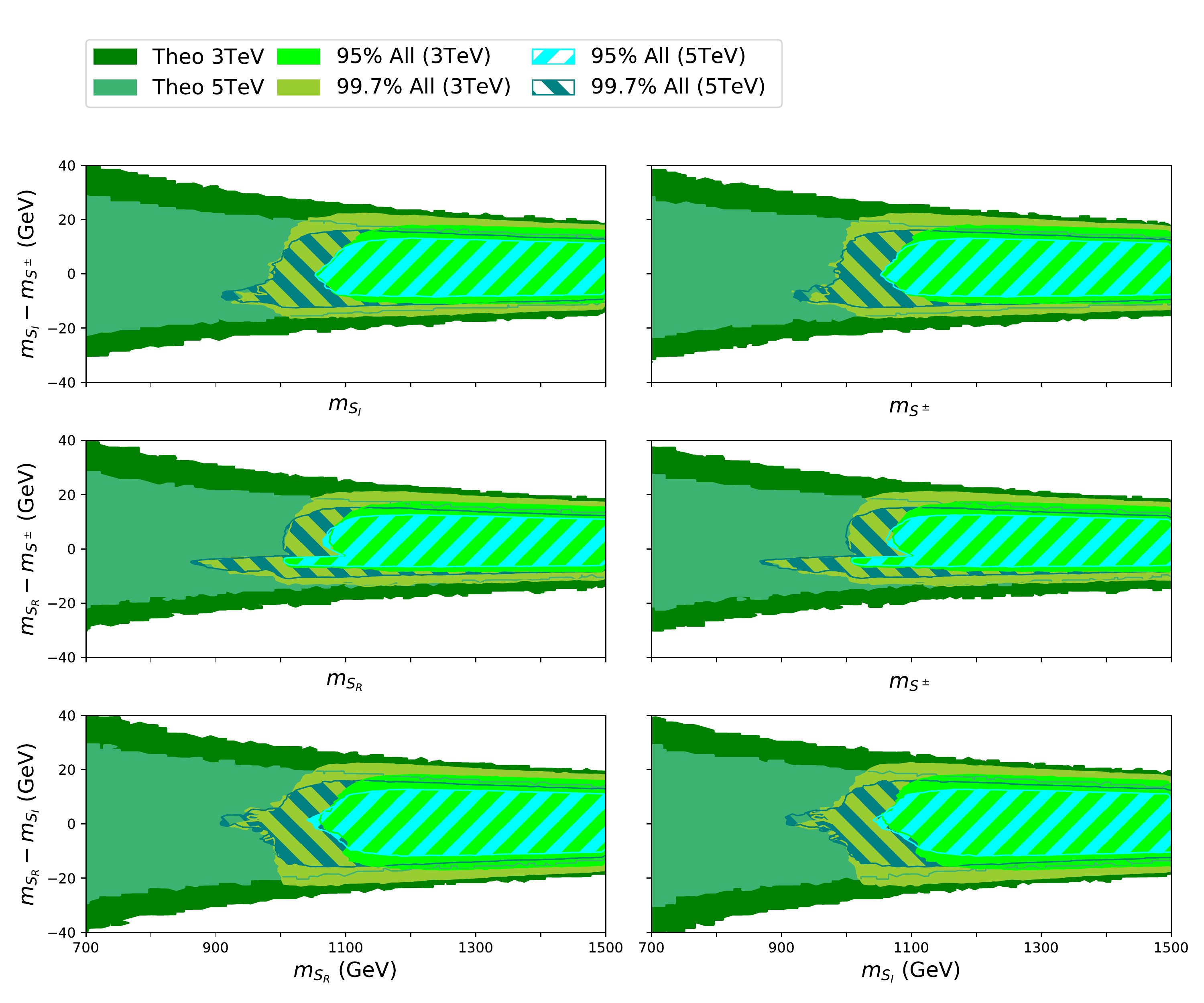}
\end{minipage}

    \caption{Left: Theoretical constraints on the mass splittings, as functions of the scalar masses. The different coloured regions correspond to imposing RG stability up to an UV mass scale of 3 or 5 TeV, and imposing perturbative unitarity conditions above 2.5 TeV or 1.5 TeV. Right: Theoretical constraints versus global fit constraints on the mass splittings.}
    \label{fig:delta_mass_teo_all}
\end{figure}

Using the theoretical considerations we are able to constrain all the parameters of the scalar potential (except for $m_S$). Some of these constraints can be translated into constraints on physical observables like the mass splittings, using Eq.~\ref{eq:msplit}. In the left part of Fig.~\ref{fig:delta_mass_teo_all} we can see the result from these fits, which are compared with the result of the global fit in the right part of the same figure. So far, the mass splittings are constrained to be smaller than 30 GeV for masses above 1 TeV within a 95\% probability.

\begin{figure}[h!]
\begin{minipage}{0.5\linewidth}
    \centering
    \includegraphics[scale=0.47]{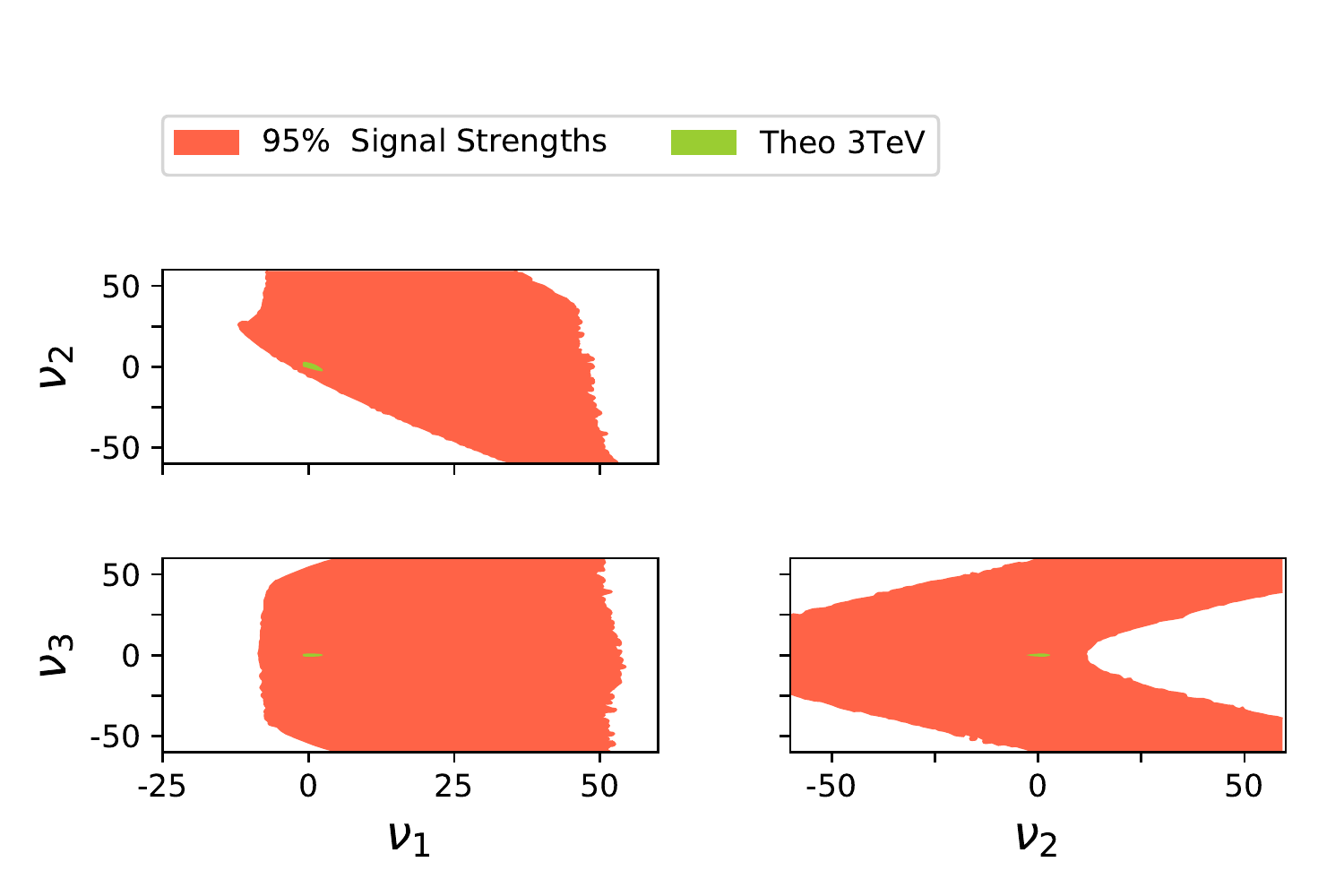}
\end{minipage}
\begin{minipage}{0.5\linewidth}
    \centering
    \includegraphics[scale=0.47]{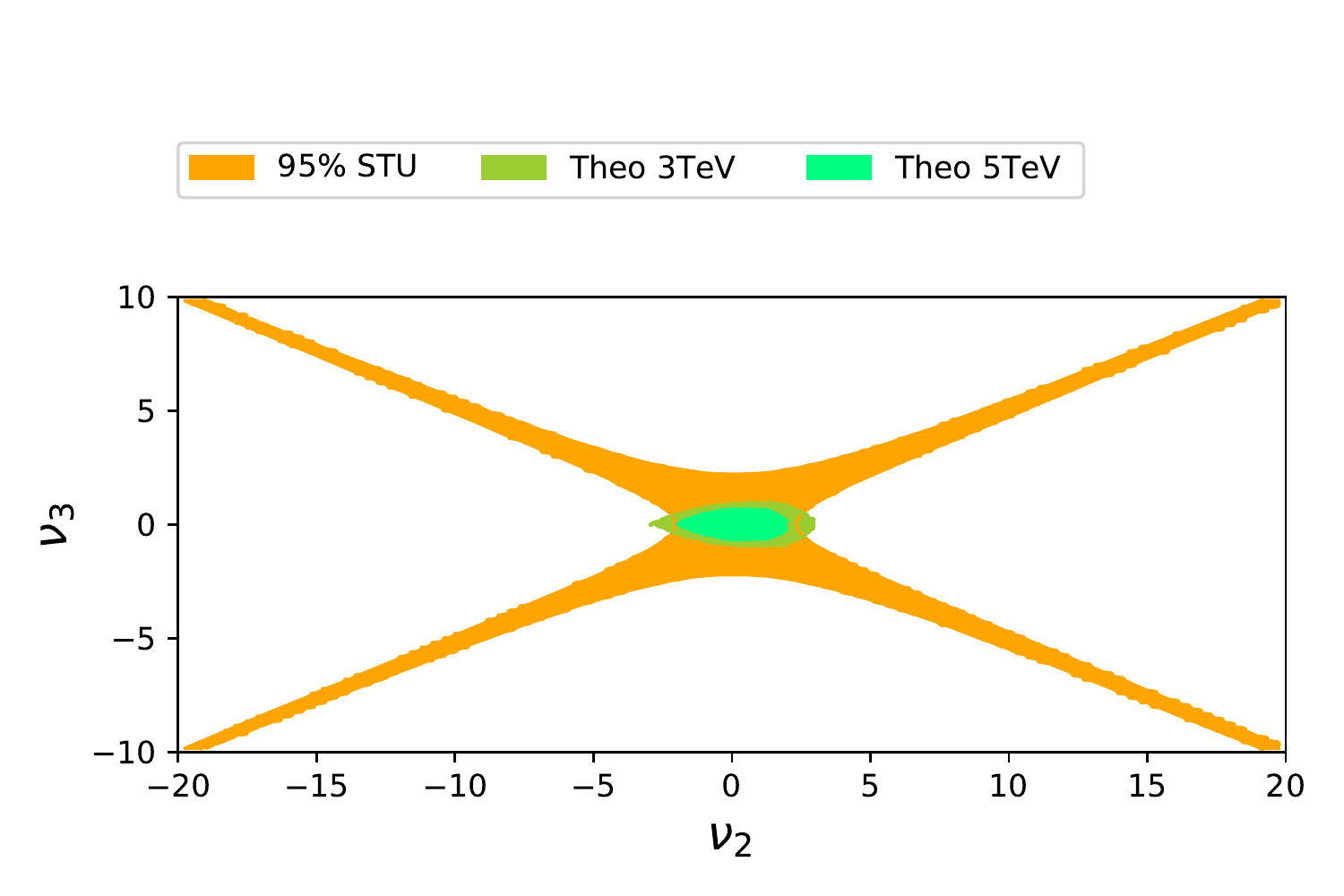}
\end{minipage}
    
    \caption{Left: Constraints obtained using HSS, compared to the theoretical ones. Right: Constraints obtained using the oblique parameters, compared to the theoretical ones.}
    \label{fig:HSS_and_STU}
\end{figure}

The parameters of the potential that generate the mass splittings, $\nu_1$, $\nu_2$ and $\nu_3$, can also be constrained using the HSS and the oblique parameters. While the HSS alone are not able to completely restrict the parameter space (see Fig.~\ref{fig:HSS_and_STU} left), the oblique parameters provide limits competitive with the theoretical ones (see Fig.~\ref{fig:HSS_and_STU} right). Furthermore, the HSS can have an important effect when combined with the oblique parameters, as we will discuss in Sec.~\ref{sec:CDFmeasurement}.  

\vspace*{-0.49 cm}

\begin{figure}[t!]
\begin{minipage}{0.5\linewidth}
    \centering
     \includegraphics[scale=0.5]{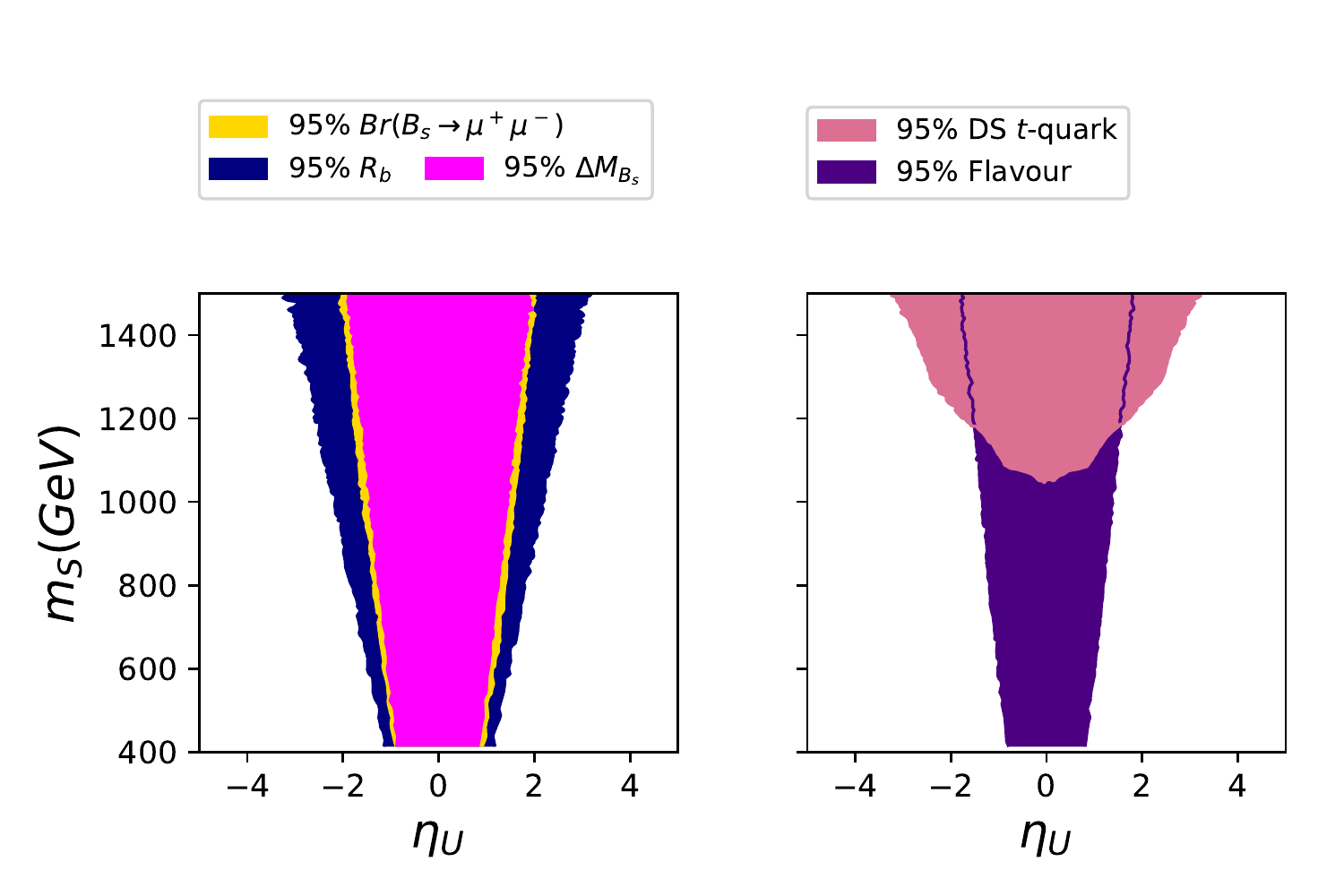}
\end{minipage}
\begin{minipage}{0.5\linewidth}
    \centering
     \includegraphics[scale=0.5]{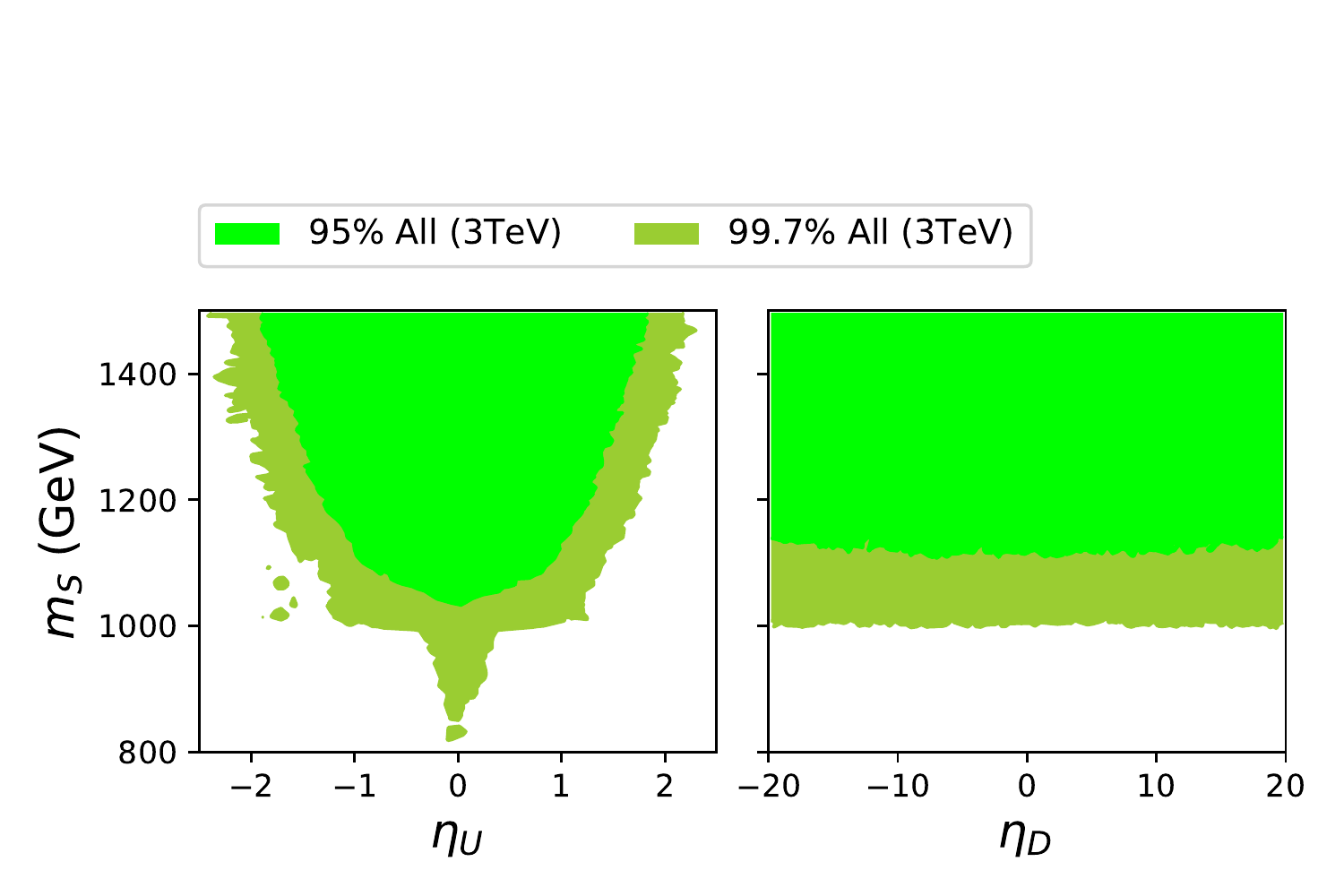}
\end{minipage}
    \caption{ Experimental constraints on the $m_S$-$\eta_U$ plane (three first plots) and on the $m_S$-$\eta_D$ plane. From left to right, first panel: allowed regions at 95\% probability obtained from $R_b$ (blue), $\rm{Br}(B_s\rightarrow \mu^+\mu^-)$ (yellow) and $\Delta M_{B_s}$ (magenta). Second panel: combined flavour constraints, compared with the limits from direct searches including only the channels that produce top quarks, at 95\% probability. Third and forth panels: Global fit constraints. }
    \label{fig:mS_Eta_planes}
\end{figure}

The flavour observables and $R_b$ will set constraints on the $m_S-\eta_U$ plane. Among them the flavour observables are the ones that have more impact (first panel of Fig.~\ref{fig:mS_Eta_planes}). From direct searches we can constrain the mass of these particles to be higher than 1 TeV and only the channels where top quarks are produced in the final state are relevant (second panel of Fig.~\ref{fig:mS_Eta_planes}). The down-type Yukawa, $\eta_D$, could not be constrained within the range considered.

\vspace*{-0.49 cm}

\section {CP-violating case: Contribution to neutron EDM}
\label{sec:EDMs}

\begin{figure}[h!]
    \centering
    \includegraphics[scale=0.41]{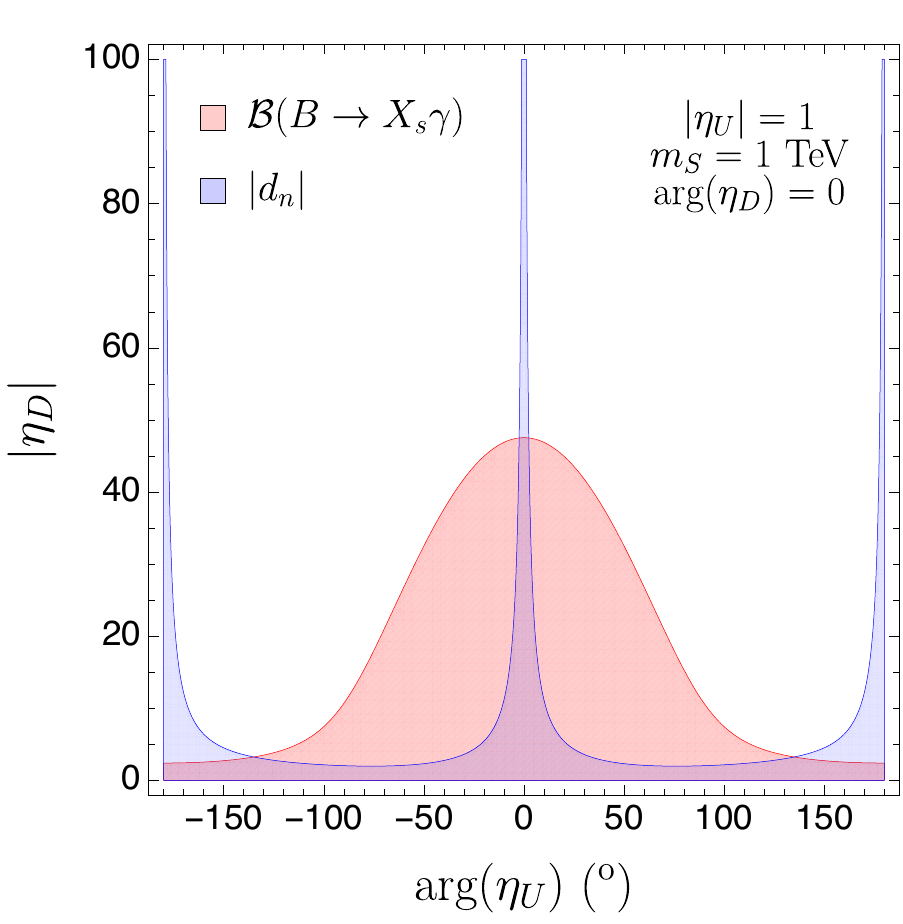}
    \includegraphics[scale=0.41]{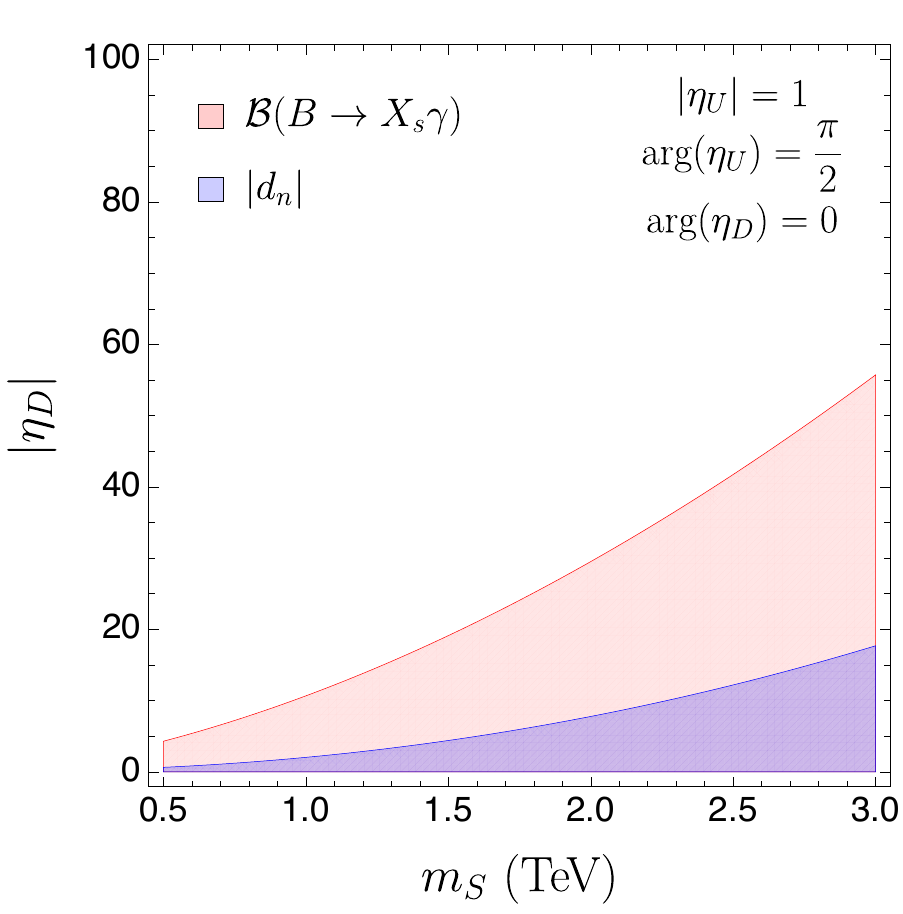}
    \includegraphics[scale=0.41]{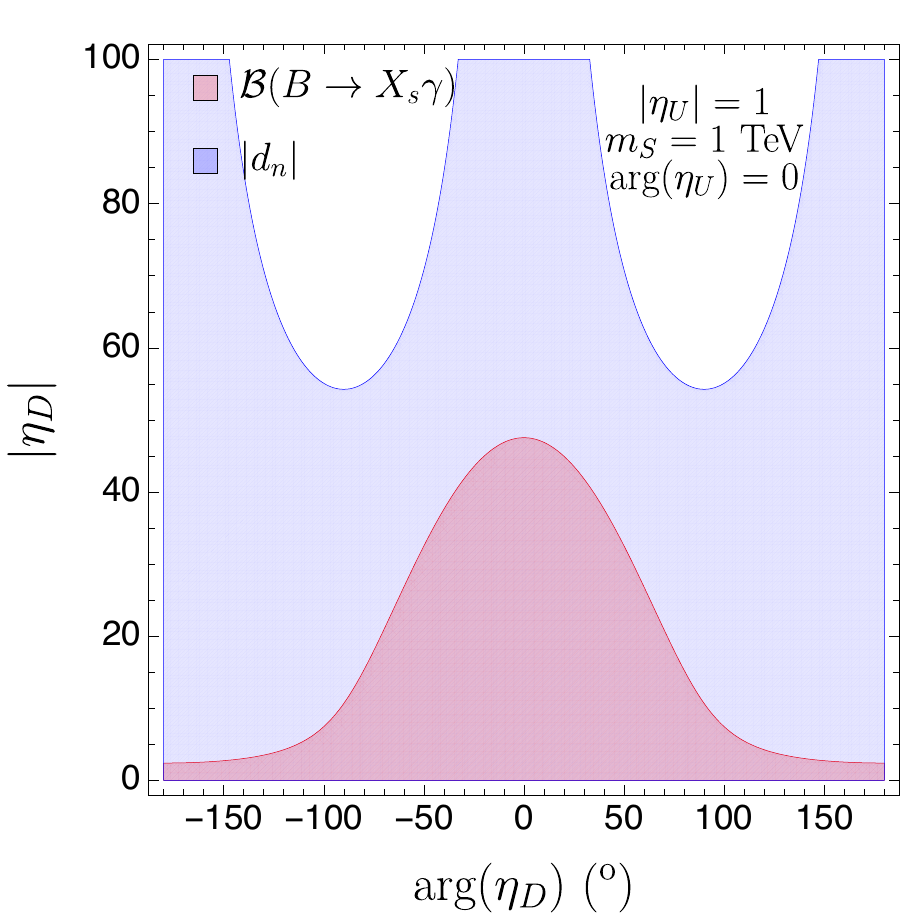}
    \includegraphics[scale=0.41]{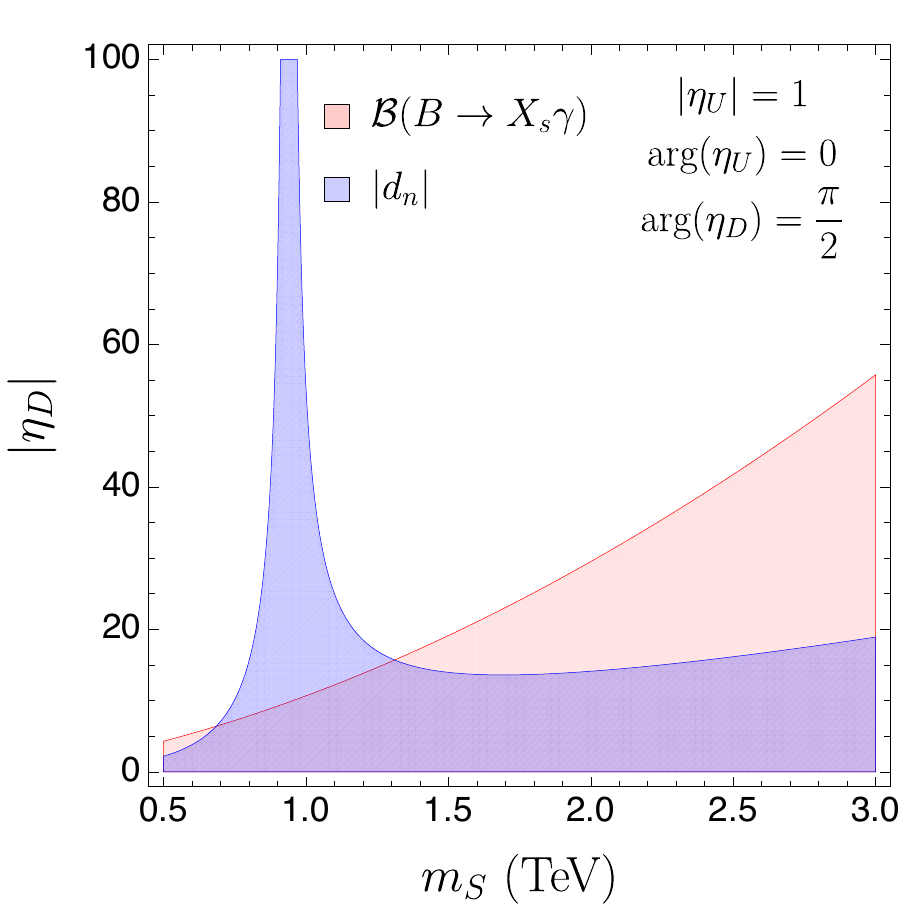}
    \caption{ Constraints on the parameter space of the MW model from the limits on the neutron EDM (blue) compared to those from $\mathcal{B}(B\rightarrow X_s\gamma)$ (red). The coloured areas represent the allowed regions of parameter space.}
    \label{fig:2D_RegionPlot_EDM}
\end{figure}

If we consider the most general case and allow for having complex phases this model will bring strong contributions to hadronic EDMs. This section is based on Ref.~\citelow{Gisbert:2021htg} where we computed the contribution of the Yukawa sector of colour-octet scalars to the EDM of the neutron. The CP-violating effective Lagrangian needed to account for the contributions of this model to the neutron EDM is given by:
\begin{equation}\label{eq:lagrangian}
        \mathcal{L}_{\rm{CPV}}\,=-\frac{i}{2}\sum_{q}^{u,d}\Big(d_q(\bar{q}\sigma^{\mu\nu}\gamma_5q)F_{\mu\nu}+\tilde{d}_q g_s(\bar{q}\sigma^{\mu\nu}\gamma_5T^a q)G^a_{\mu\nu}\Big)+ \frac{C_w}{6}g_sf^{a b c}\epsilon^{\mu\nu\lambda\sigma}G^{a}_{\mu\rho}G^{b \rho}_{\nu}\,G^{c}_{\lambda\sigma}
\end{equation}
Therefore, we needed to compute the contribution of the colour-octet scalars to the EDM ($d_q$) and chromo-EDM (CEDM) ($\tilde{d}_q$) of the light quarks and to the Weinberg operator ($C_w$). Although there is already a 1-loop contribution to the (C)EDM of light quarks, the leading contribution will be given by the well known 2-loop Barr-Zee diagrams, since the 1-loop contribution is strongly suppressed by the masses of the light quarks. The first contribution to the Weinberg operator also appears at 2-loops.

Once the contribution to the neutron EDM has been computed, we can compare its effect with the one of other flavour observables.
Indeed, the most constraining observable for the complex phases is coming from the rare inclusive $B$-decays, $\mathcal{B}(B\rightarrow X_s\gamma)$, calculated in Ref.~\citelow{Cheng:2015lsa}. As can be seen in Fig.~\ref{fig:2D_RegionPlot_EDM} the neutron EDM turns out to be extremely relevant, bringing very stringent constraints and with a great complementarity with the flavour observables.

\section{Impact on the recent CDF measurement of the W-boson mass}
\label{sec:CDFmeasurement}

\vspace*{-0.2 cm}
\begin{figure}[h!]
\begin{minipage}{0.5\linewidth}
Given the huge attention that the new measurement of the mass of the $W$-boson by CDF \cite{CDF:2022hxs} is receiving, we show in Fig.~\ref{fig:CDF} the favoured regions of a fit using the new values of the oblique parameters, extracted in Ref.~\citelow{deBlas:2022hdk} using this measurement.
There are already some works that study this new measurement in the context of colour-octet scalars \cite{Carpenter:2022oyg,Gisbert:2022lao}. Here we also show the results of combining the oblique parameters with the HSS and those from theoretical assumptions, which have not been discussed before. So far, adding the HSS we are able to eliminate all the blind directions. When, in addition, we include the theoretical assumptions (imposed up to 3 TeV) the favoured regions move to the centre indicating the existence of a strong tension between these two fits. Indeed, the region favoured by the fit that only includes the oblique parameters is incompatible with the region allowed by the theoretical constraints. 
\end{minipage}\hspace*{0.2 cm}\begin{minipage}{0.5\linewidth}
    \centering
    \includegraphics[scale=0.43]{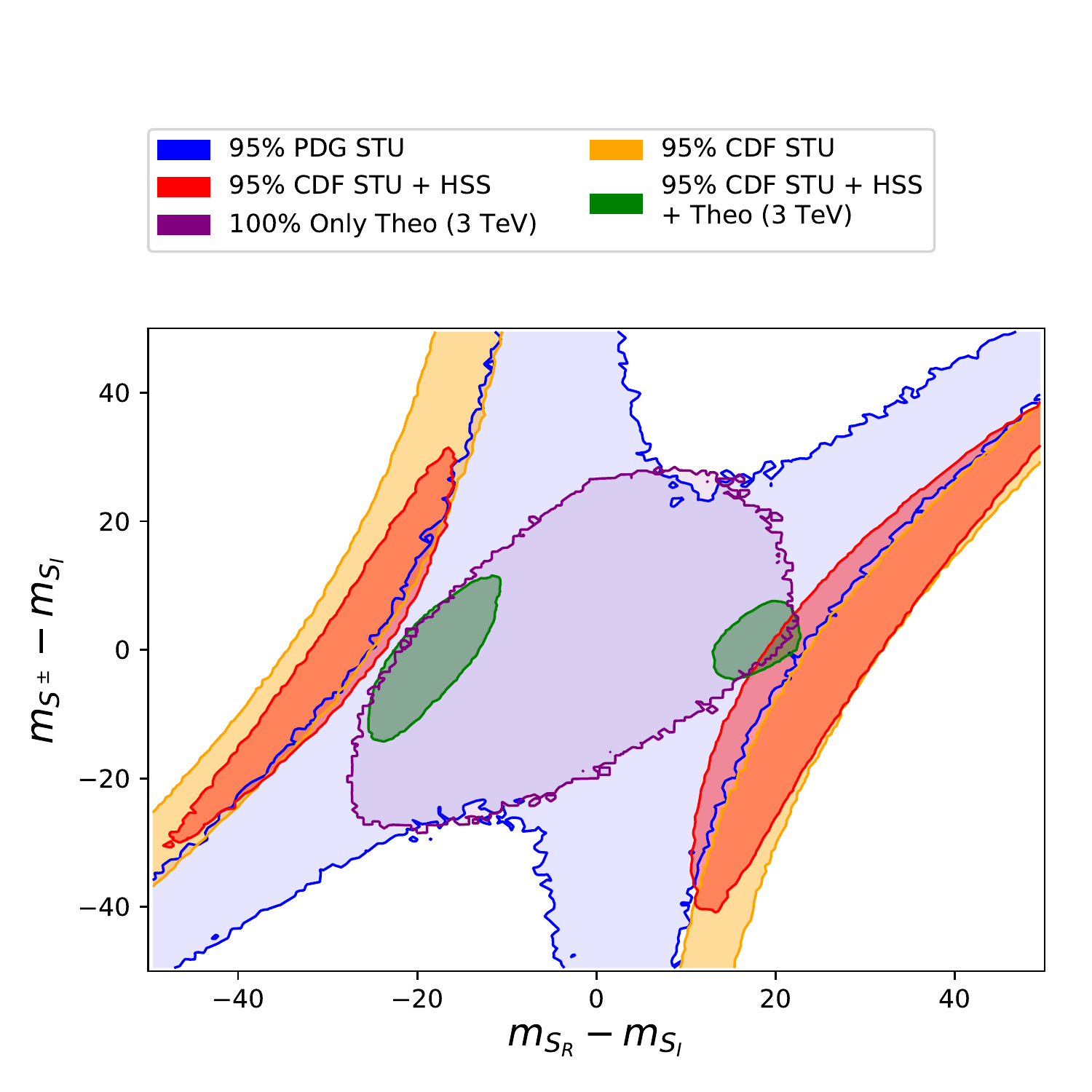}
    \caption{Constraints on the mass splitting using the oblique parameters of the PDG (blue), the ones including the CDF measurement (orange), adding to these ones the HSS (red), adding also the theoretical assumptions up to 3 TeV (green) and those including only the theoretical assumptions (purple). $m_{S^\pm}$ is set to 1 TeV.}
    \label{fig:CDF}
\end{minipage}
\end{figure}

\vspace*{-0.2 cm}

\section{Conclusion}

We have presented the most stringent limits that can be set in the parameter space of the MW model. A global fit of the CP-conserving case is performed, as well as an analysis of the impact of the EDM of the neutron on the imaginary phases. Indeed, the EDM turns out to be very restrictive, and competitive with other flavour observables. Finally, the impact of the new measurement of the $W$-boson mass has been also analysed. This new measurement pushes the masses of the colour-octet particles to be non degenerate. However, the theoretical constraints are not compatible with the region preferred by the fit including only the oblique parameters, disfavouring this model as an explanation for the new $W$-boson mass measurement.

\section*{Acknowledgments}

We thank L. Silvestrini for sharing the details of the SM fit and interesting discussions. 
This work has been supported by MCIN/AEI/10.13039/501100011033, Grant No. PID2020-114473GB-I00,
by the Generalitat Valenciana, Grant No. Prometeo/2021/071, and by the Italian Ministry of Research (MUR) under
the grant PRIN20172LNEEZ.

\section*{References}

\bibliography{references}

\end{document}